\documentclass{article}
\usepackage{amssymb}
\usepackage[a4paper]{geometry}
\usepackage{amsmath}

\input{tcilatex}
\begin{document}

\title{Solutions to the time-dependent Schrodinger equation in the
continuous spectrum case }
\author{M. Maamache and Y. Saadi \\
\\
\textit{Laboratoire de Physique Quantique et Syst\`{e}mes Dynamiques,}\\
\textit{{Facult\'{e} des Sciences,Universit\'{e} Ferhat Abbas de S\'{e}tif},
S\'{e}tif 19000, Algeria}}
\date{}
\maketitle

\begin{abstract}
We generalize the Lewis-Riesenfeld technique of solving the time-dependent
Schrodinger equation to cases where the invariant has continuous
eigenvalues. An explicit formula for a generalized Lewis-Riesenfeld phase is
derived in terms of the eigenstates of the invariant. As an illustration the
generalized phase is calculated for a particle in a time-dependent linear
potential.

\end{abstract}

The study of time dependent quantum systems has attracted considerable
interest in the litterature. The origin of this development was no doubt the
discovery of an exact invariant by Lewis and Riesenfeld \cite{1} . The work
of Lewis and Riesenfeld and others assumes that the eigenvalue spectrum for
the invariant $I$ \ is discrete. Let us recall that the general method to
introduce the Lewis and Riesenfeld theory, valid whatever the time
dependence of the parameters, considers invariant operators. For a system
specified by a time-dependent Hamiltonian $H\left( \vec{X}\left( t\right)
\right) $, and a corresponding evolution operator $U\left( t\right) $, an
invariant is an operator $I(t)$ such that

\bigskip 
\begin{equation}
\frac{dI}{dt}=\frac{\partial I}{\partial t}+\frac{1}{i\hbar }\left[ I,H%
\right] =0.  \label{1}
\end{equation}

\bigskip or

\begin{equation}
I(t)=U\left( t\right) I(0)\text{ }U^{-1}\left( t\right) .  \label{2}
\end{equation}%
It possesses a remarkable property that any eigenstate of $I(0)$ evolves
into an eigenstate of $I(t)$. Then, if the set of reference eigenstates $%
\{\mid \phi _{n}(t)\rangle \}$ for the operator $I(t)$ are continuous with
respect to $t$ (all eigensates are associated with the same time-independent
eigenvalue $\varepsilon _{n}$), the corresponding global phases $\theta
_{n}(t)$ are defined by the relation associated to the wave functions $%
\left\vert \psi _{n}(t)\right\rangle $:

\begin{equation}
\left\vert \psi _{n}(t)\right\rangle =U\left( t\right) \mid \phi
_{n}(0)\rangle \text{ }=e^{i\theta _{n}(t)}\mid \phi _{n}(t)\rangle .
\label{3}
\end{equation}%
It follows from the Schr\"{o}dinger equation for $\left\vert \psi
_{n}(t)\right\rangle $

\begin{equation}
i\hbar \frac{\partial }{\partial t}\left\vert \psi _{n}(t)\right\rangle
=H(t)\psi _{n}(t)\left\vert \psi _{n}(t)\right\rangle ,  \label{4}
\end{equation}%
that $\theta _{n}(t)$ satisfies the relation

\begin{equation}
\hbar \frac{d}{dt}\theta _{n}(t)=\langle \phi _{n}(t)\mid i\hbar \frac{%
\partial }{\partial t}-H\mid \phi _{n}(t)\rangle .  \label{5}
\end{equation}

One way to describe the exact quantum evolution of \ Lewis and Riesenfeld is
to introduce the concept of elementary projectors on an eigenstate $\mid
\phi _{n}(t)\rangle $ of the invariant operator $I(t)$%
\begin{equation}
P_{n}(t)=\mid \phi _{n}(t)\rangle \langle \phi _{n}(t)\mid .  \label{6}
\end{equation}%
It is easy to verify that each projector $P_{n}\left( t\right) $ is
therefore a constant of the motion i.e., $P_{n}\left( t\right) =U\left(
t\right) P_{n}\left( 0\right) U^{+}\left( t\right) .$

We can state that the exact evolution described by equations (3) and (5) can
be formally written in the following form

\begin{equation}
\forall t:\ U\left( t\right) P_{n}\left( 0\right) =P_{n}\left( t\right)
U\left( t\right) .  \label{7}
\end{equation}%
Notice that if, initially, the system is in the eigenstate $\left\vert \phi
_{n}\left( 0\right) \right\rangle $ so that $I\left( 0\right) \left\vert
\phi _{n}\left( 0\right) \right\rangle =\varepsilon _{n}\left\vert \phi
_{n}\left( 0\right) \right\rangle $, then $P_{n}\left( 0\right) \left\vert
\phi _{n}\left( 0\right) \right\rangle =\left\vert \phi _{n}\left( 0\right)
\right\rangle $ and (\ref{7}) gives%
\begin{equation}
U\left( t\right) \left\vert \phi _{n}\left( 0\right) \right\rangle
=P_{n}\left( t\right) U\left( t\right) \left\vert \phi _{n}\left( 0\right)
\right\rangle .  \label{8}
\end{equation}

In general the spectrum of $I$ possess both discret and continuous
eigenvalues. The Lewis and Riesenfeld theory in a continuous spectrum was
raised for the first time by Hartley and Ray \cite{2} where they extend the
Lewis and Riesenfeld theory for a general Ermakov system to the continuous
spectra like an anstaz and looks at the eigenfunctions in a continuous
spectrum $\left\vert \phi \left( k;t\right) \right\rangle $ of the invariant
operator $I(t)$ and the solution $\left\vert \psi (k,t)\right\rangle $of the
Schrodinger equation in the form $e^{i\theta _{k}(t)}\left\vert \phi \left(
k;t\right) \right\rangle .$The limitation of the Hartley-Ray approach \cite%
{2} is that, in general, there is no explicit formula of Lewis and
Riesenfeld phase. Later Gao et al \cite{3} calculate the Schr\"{o}dinger
solutions for a continuous spectrum using  the path integral technique. In
some papers \cite{3} \cite{4} the overall phase factor $\theta _{k}(t)$ ,
interpreted in the spirit of the original investigation of Lewis and
Riesenfeld, is actually obtained through the relation $\hbar \frac{d}{dt}%
\theta _{k}(t)=\langle \phi (k;t)\mid i\hbar \frac{\partial }{\partial t}%
-H\mid \phi (k;t)\rangle .$This procedure clearly parallel to that for the
discrete spectrum is not founded. The reason is that the calculation of the
expectation value $(i\hbar \frac{\partial }{\partial t}-H)$ with respect to
the eigenfunctions $\left\vert \phi \left( k;t\right) \right\rangle $
analyzed trough examples in \cite{4} lead to errenous results.

In the case of continuous spectrum we cannot numerate eigenvalues and
eigenfunctions, they are characterised by the value of the physical quantity
in the corresponding state. Althoug the eigenfunctions $\left\vert \phi
\left( k;t\right) \right\rangle $of the invariant operators with continuous
spectra cannot be normalised in the usual manner as is done for the
functions of discret spectra, one can construct with the $\left\vert \phi
\left( k;t\right) \right\rangle $ new quantities - theWeyl's\ \textit{%
eigendifferentials (wave packets)- }\cite{5} \cite{6} which possess the
properties of the eigenfunction of discrete spectrum. The eigendifferentials
are defined by the equation%
\begin{equation}
\left\vert \delta \phi \left( k;t\right) \right\rangle =\overset{k+\delta k}{%
\underset{k}{\int }}\left\vert \phi \left( k^{\prime };t\right)
\right\rangle dk^{\prime }.  \label{9}
\end{equation}%
They divide up the continuous spectrum of the eigenvalues into finite but
sufficiently small discrete regions of size $\delta k$ .

The eigendifferential (\ref{9}) is a special wave packet which has only a
finite extension in space; hence, it vanishes at infinity and therefore can
be seen in analogy to bound states. Furthermore, because the $\delta \varphi 
$ have finite spatial extension, they can be normalized. Then in the limit $%
\delta k\rightarrow 0$,\ a meaningful normalization of the function $\varphi 
$ themselves follows: the normalization on $\delta $ functions.

For $\delta k$, a small connected range of value of the parameter $k$ (this
corresponds to a group of "neighboring" states), the operator%
\begin{equation}
\delta P\left( k;t\right) =\underset{k}{\overset{k+\delta k}{\dint }}%
\left\vert \phi \left( k^{\prime };t\right) \right\rangle \left\langle \phi
\left( k^{\prime };t\right) \right\vert dk^{\prime }  \label{10}
\end{equation}%
represents the projector (the differential projection operator \cite{5}\cite%
{6}) onto those states contained in the interval and characterized by the
values of the parameter $k$\ within the range of values $\delta k$. The
action of $\delta P\left( k;t\right) $ on a wavefunction $\left\vert \psi
\left( t\right) \right\rangle $ is defined by%
\begin{equation}
\delta P\left( k;t\right) \left\vert \psi \left( t\right) \right\rangle =%
\underset{k}{\overset{k+\delta k}{\dint }}C\left( k^{\prime };t\right)
\left\vert \phi \left( k^{\prime };t\right) \right\rangle dk^{\prime }.
\label{11}
\end{equation}

The application of the differential projection operator $\delta P\left(
k;t\right) $ causes thus the projection of the wavefunction onto the domain
of states $\phi \left( k;t\right) $ which is characterized by $k$\ values
within the $\delta k$ interval .

In this letter, we present a straightforward, yet rigorous, proof of the
exact quantum evolution for systems whose invariant has a completely
continuous spectrum supposed to be  non-degenerated for reasons of
simplicity. Example of a particle in a time-dependent linear potential is
worked out for illustration. The case of both disrete and continuous
eigenvalues can be obtained by superposition. Before proceeding further, we
give the statement of the exact quantum evolution.

Given a physical system with a time-dependent Hamiltonian $H\left( t\right) $%
, it is possible to build an invariant operator $I(t)$ verifying (\ref{1}),
such that its eigenvalues $\varepsilon _{k}$ are purely continuous and
constants

\begin{equation}
I(t)\left\vert \phi \left( k;t\right) \right\rangle =\varepsilon
_{k}\left\vert \phi \left( k;t\right) \right\rangle .  \label{12}
\end{equation}%
Let us call $U\left( t\right) $ the evolution operator associated to the
time-dependent Hamiltonian $H\left( t\right) $. The evolution of the system
obeys to the Schr\"{o}dinger equation 
\begin{equation}
i\hbar \frac{\partial }{\partial t}\left\vert \psi (t)\right\rangle =H\left(
t\right) \left\vert \psi (t)\right\rangle ,  \label{13}
\end{equation}%
under these conditions it is possible to state the exact quantum evolution:

"\textit{If the quantum system with time-dependent Hamiltonian whose
invariant has a completely continuous spectrum supposed non-degenerated is
initially in an eigenstate }$\left\vert \phi \left( k,0\right) \right\rangle 
$\textit{\ of }$I\left( 0\right) $\textit{\ then the state of the system at
any time }$t$\textit{\ will remain in the subspace generated by the
eigenstates }$\left\vert \phi \left( k;t\right) \right\rangle $\textit{\ of }%
$I(t)$ \textit{pertaining to the interval }$\left[ k,k+\delta k\right] $".

In others words, the exact evolution can be formally written in terms of the
evolution operator as%
\begin{equation}
\forall k,\forall t:\text{ \ }U\left( t\right) \delta P\left( k;0\right)
=\delta P\left( k;t\right) U\left( t\right) .  \label{14}
\end{equation}

\bigskip The proof of this last equation is straightforward. One has to take
the partial derivative of $\delta P\left( k;t\right) =\underset{k}{\overset{%
k+\delta k}{\dint }}\left\vert \phi \left( k^{\prime };t\right)
\right\rangle \left\langle \phi \left( k^{\prime };t\right) \right\vert
dk^{\prime }$ with respect to the time $t$, taking into acount Eqs. (1) and
(12), one will get then: 
\begin{equation}
\frac{\partial \delta P\left( k;t\right) }{\partial t}+\frac{1}{i\hbar }%
\left[ \delta P\left( k;t\right) ,H\right] =0,  \label{15}
\end{equation}%
which has a formal solution $\delta P\left( k;t\right) =U\left( t\right)
\delta P\left( k;0\right) U^{+}\left( t\right) .$

Notice that if, initially, the system is in the state $\left\vert \psi
(k,0)\right\rangle =\left\vert \phi \left( k,0\right) \right\rangle $, then (%
\ref{14}) implies that

\bigskip 
\begin{equation}
\delta P\left( k^{\prime };t\right) U\left( t\right) \left\vert \psi \left(
k,0\right) \right\rangle =U\left( t\right) \delta P\left( k^{\prime
};0\right) \left\vert \psi (k,0)\right\rangle ,  \label{16}
\end{equation}%
expanding an arbitrary state vector $\left\vert \psi \left( t\right)
\right\rangle $ on the basis of the instantaneous quasi-eigenfunction of $%
I(t)$ and using (11) we obtain%
\begin{equation}
\left\vert \psi \left( k,t\right) \right\rangle =\underset{k}{\overset{%
k+\delta k}{\dint }}C\left( k^{\prime };t\right) \left\vert \phi \left(
k^{\prime };t\right) \right\rangle dk^{\prime };\text{ \ \ \ }\forall
k^{\prime }\in \left[ k,k+\delta k\right] ,  \label{17}
\end{equation}%
we see that the state $\left\vert \psi \left( k,t\right) \right\rangle $
belongs to the subspace generated by the states $\left\vert \phi \left(
k;t\right) \right\rangle $ pertaining to the interval $\left[ k,k+\delta k%
\right] .$

Inserting (\ref{17}) in the Schr\"{o}dinger equation (\ref{13}), lead to

\begin{equation}
\overset{k+\delta k}{\underset{k}{\int }}i\hbar \frac{\partial }{\partial t}%
C\left( k^{\prime };t\right) \left\vert \phi \left( k^{\prime };t\right)
\right\rangle dk^{\prime }+\overset{k+\delta k}{\underset{k}{\int }}i\hbar
C\left( k^{\prime };t\right) \frac{\partial }{\partial t}\left\vert \phi
\left( k^{\prime };t\right) \right\rangle dk^{\prime }=\overset{k+\delta k}{%
\underset{k}{\int }}C\left( k^{\prime };t\right) H\left( k^{\prime
};t\right) \left\vert \varphi \left( k^{\prime };t\right) \right\rangle
dk^{\prime }.  \label{18}
\end{equation}%
We multiply Eq. (\ref{18}) by the bra of the eigendifferential (\ref{9})
introduced earlier%
\begin{equation}
\left\langle \delta \phi \left( \gamma ;t\right) \right\vert =\overset{%
\gamma +\delta \gamma }{\underset{\gamma }{\int }}\left\langle \phi \left(
\eta ;t\right) \right\vert d\eta   \label{19}
\end{equation}%
this yields%
\begin{equation}
\overset{k+\delta k}{\underset{k}{\int }}i\hbar \frac{\partial }{\partial t}%
C\left( k^{\prime };t\right) \left\vert \phi \left( k^{\prime };t\right)
\right\rangle dk^{\prime }=\overset{k+\delta k}{\underset{k}{\int }}C\left(
k^{\prime };t\right) \left\langle \delta \phi \left( \gamma ;t\right)
\right\vert H(t)-i\hbar \frac{\partial }{\partial t}\left\vert \phi \left(
k^{\prime };t\right) \right\rangle dk^{\prime }  \label{20}
\end{equation}%
Since $k$ can sweep all the possible values and the intervals $\delta k$
should be small $\left( \delta k\rightarrow 0\right) $, the equality (\ref%
{20}) between integrals implies the equality between integrands, hence%
\begin{equation}
i\hbar \frac{\partial }{\partial t}C\left( k^{\prime };t\right) =C\left(
k^{\prime };t\right) \left[ \left\langle \delta \phi \left( k;t\right)
\right\vert H(t)-i\hbar \frac{\partial }{\partial t}\left\vert \phi \left(
k^{\prime };t\right) \right\rangle \right] ;\ k^{\prime }\in \left[
k,k+\delta k\right] .  \label{21}
\end{equation}%
This equation is easily integrated and gives:%
\begin{equation}
C\left( k^{\prime };t\right) =\delta \left( k^{\prime }-k\right) \exp \left[
-i\underset{0}{\overset{t}{\int }}\left( \left\langle \delta \phi \left(
k;t^{\prime }\right) \right\vert \frac{1}{\hbar }H\left( t^{\prime }\right) -%
\frac{\partial }{\partial t^{\prime }}\left\vert \phi \left( k^{\prime
};t^{\prime }\right) \right\rangle \right) dt^{\prime }\right] ;\ k^{\prime
}\in \left[ k,k+\delta k\right] ,  \label{22}
\end{equation}%
hence%
\begin{equation}
\left\vert \psi \left( k,t\right) \right\rangle =\exp \left\{ \frac{i}{\hbar 
}\theta _{k}(t)\right\} \left\vert \phi \left( k;t\right) \right\rangle ,
\label{23}
\end{equation}%
where $\theta _{k}(t)$ is the global phase given by%
\begin{equation}
\theta _{k}(t)=\underset{0}{\overset{t}{\int }}\left\langle \delta \phi
\left( k;t^{\prime }\right) \right\vert i\frac{\partial }{\partial t^{\prime
}}-\frac{1}{\hbar }H\left( t^{\prime }\right) \left\vert \phi \left(
k;t^{\prime }\right) \right\rangle dt^{\prime }.  \label{24}
\end{equation}%
This explicit formula of the phase could not be made in the Hartley-Ray
approach \cite{2} as mentioned earlier.

As the interval $\left[ k,k+\delta k\right] $ is located inside of the
interval $\left[ -\infty ,+\infty \right] $, we can write the generalized
phase in the following practical form

\begin{equation}
\theta _{k}(t)=\underset{0}{\overset{t}{\int }}\underset{-\infty }{\overset{%
+\infty }{\int }}\left\langle \phi \left( k^{\prime };t^{\prime }\right)
\right\vert i\frac{\partial }{\partial t^{\prime }}-\frac{1}{\hbar }H\left(
t^{\prime }\right) \left\vert \phi \left( k;t^{\prime }\right) \right\rangle
dt^{\prime }dk^{\prime },  \label{25}
\end{equation}%
which embodies the central result of this paper.

To illustrate this theory, let us calculate this phase for a particle moving
in a time dependent linear potential

\begin{equation}
H\left( t\right) =\frac{1}{2m}p^{2}+f\left( t\right) x,  \label{26}
\end{equation}%
where $f(t)$ is a time-dependent function. We look for the invariant of the
form%
\begin{equation}
I\left( t\right) =a\left( t\right) p^{2}+b\left( t\right) p+c\left( t\right)
x+d\left( t\right) .  \label{27}
\end{equation}

The invariant equation (1) is satisfied if the time-dependent coefficients
are such that

\begin{equation}
a=a_{0},  \label{28}
\end{equation}%
\begin{equation}
b=2a_{0}\int_{0}^{t}f\ dt^{\prime }-c_{0}\int_{0}^{t}\frac{1}{m}dt^{\prime
}+b_{0},  \label{29}
\end{equation}%
\begin{equation}
c=c_{0},  \label{30}
\end{equation}%
\begin{equation}
d=2a_{0}\int_{0}^{t}f\ \int_{0}^{t^{\prime }}f\ dt^{\prime
}dt"-c_{0}\int_{0}^{t}f\ \int_{0}^{t^{\prime }}\frac{1}{m}dt^{\prime
}dt"+b_{0}\int_{0}^{t}f\ dt^{\prime }+d_{0},  \label{31}
\end{equation}%
where $a_{0}$, $b_{0}$, $c_{0}$ and $d_{0}$ are arbitrary real constants. We
can choose $a_{0}=1$ and $d_{0}=0$ without loss of generalities. The
eigenstates of $I(t)$ corresponding to time-independent eigenvalues $k$ are
the solutions of the equation

\begin{equation}
\left[ -\hbar ^{2}\frac{\partial ^{2}}{\partial x^{2}}-i\hbar b\frac{%
\partial }{\partial x}+c_{0}x+d\right] \ \phi _{k}\left( x,t\right) =k\phi
_{k}\left( x,t\right) .  \label{32}
\end{equation}

The key point of our analysis is to perform the time-dependent unitary
transformation such that 
\begin{equation}
\Phi _{k}(x)=\Xi (t)\phi _{k}(x,t),  \label{33}
\end{equation}%
where a time-dependent unitary operator $\Xi (t)$ is given by 
\begin{equation}
\Xi (t)=\exp \left( i\frac{1}{\hbar c_{0}}\left( \frac{b^{2}}{4}-d\right)
p\right) \times \exp \left( i\frac{b}{2\hbar }x\right) .  \label{34}
\end{equation}%
It can be easily shown that, under this transformation, the coordinate and
momentum operators change according to

\bigskip 
\begin{equation}
x~\longrightarrow ~\Xi (t)x\Xi (t)^{+}=x+\frac{1}{c_{0}}\left( \frac{b^{2}}{4%
}-d\right) ,  \label{35}
\end{equation}%
\begin{equation}
p~\longrightarrow ~\Xi (t)p\Xi (t)^{+}=p-\frac{b}{2}.  \label{36}
\end{equation}

\bigskip An important property of the transformation $\exp \left( -i\frac{1}{%
\hbar c_{0}}\left( \frac{b^{2}}{4}-d\right) p\right) $, the action of which
on a wave function in the $x$ representation reads%
\begin{equation}
\exp \left( -i\frac{1}{\hbar c_{0}}\left( \frac{b^{2}}{4}-d\right) p\right)
\digamma (x,t)=\digamma \left[ x-\frac{1}{c_{0}}\left( \frac{b^{2}}{4}%
-d\right) ,t\right] .  \label{37}
\end{equation}

Hence, the operator $I$ changes into time-independent operator $I_{0}=\Xi $ $%
I$ $\Xi ^{+}=p^{2}+c_{0}x$. In other words, the eigenvalue equation (32) for
the transformed invariant operator can be simply represented in the form of
Airy equation 
\begin{equation}
\left[ \frac{\partial ^{2}}{\partial Z^{2}}-Z\right] \ \Phi _{k}(Z)=0,
\label{38}
\end{equation}%
where we have introduced a new variable $Z$ related to $x$ through the
relation $Z=\left( \frac{c_{0}}{\hbar ^{2}}\right) ^{\frac{1}{3}}(x-\frac{k}{%
c_{0}}).$The solution of the fundamental one dimensional ordinary
second-order differential equation (38) is well-known 
\begin{equation}
\ \Phi _{k}(x)=\frac{1}{2\pi }\left( \frac{1}{c_{0}\hbar ^{4}}\right) ^{%
\frac{1}{6}}\ Ai\left( \left( \frac{c_{0}}{\hbar ^{2}}\right) ^{\frac{1}{3}}%
\left[ x-\frac{k}{c_{0}}\right] \right) ,  \label{39}
\end{equation}

$Ai\left( x\right) =\underset{-\infty }{\overset{+\infty }{\int }}%
e^{i\lambda x}e^{i\frac{\lambda ^{3}}{3}}d\lambda $ being the integral
representation of the Airy function. It is easy to verify that $\langle \Phi
_{k^{\prime }}\left\vert \Phi _{k}\right\rangle =\delta \left( k-k^{\prime
}\right) .$

Reversing the procedure above, we can obtain

\begin{equation}
\phi _{k}\left( x,t\right) =\ \frac{1}{2\pi }\left( \frac{1}{c_{0}\hbar ^{4}}%
\right) ^{\frac{1}{6}}\exp \left[ -i\frac{b}{2\hbar }x\right] Ai\left(
\left( \frac{c_{0}}{\hbar ^{2}}\right) ^{\frac{1}{3}}\left[ x-\frac{1}{c_{0}}%
\left( \frac{b^{2}}{4}+k-d\right) \right] \right) .  \label{40}
\end{equation}

\bigskip There remains the problem of finding the phases $\theta _{k}(t)$
which satisfy (25). Carrying out the unitary transformation $\Xi $ the right
-hand side of \ Eq. (25) becomes

\bigskip 
\begin{eqnarray}
\left\langle \phi _{k^{\prime }}\left\vert i\hbar \frac{\partial }{\partial t%
}-H\left( t\right) \right\vert \phi _{k}\right\rangle  &=&-\frac{1}{2m}%
\left\langle \Phi _{k^{\prime }}\left\vert \left[ I_{0}+\frac{b^{2}}{2}-d%
\right] \right\vert \Phi _{k}\right\rangle   \notag \\
&=&-\frac{1}{2m}\left[ k+\frac{b^{2}}{2}-d\right] \delta \left( k-k^{\prime
}\right) ,  \label{41}
\end{eqnarray}%
where we have used $I_{0}\left\vert \Phi _{k}\right\rangle =k\left\vert \Phi
_{k}\right\rangle $ and the $\delta $ Dirac normalisation of $\Phi _{k}$.
Note that if $k=k^{\prime }$ the latter Eq. (41) leads  to an unfounded
result (equal to infinity) as was mentioned earlier. Hence the phase (25) is
found to be

\begin{equation}
\theta _{k}(t)-\theta _{k}(0)=-\frac{1}{2m}\underset{0}{\overset{t}{\int }}%
\left[ k+\frac{b^{2}}{2}-d\right] dt^{\prime }.  \label{42}
\end{equation}

\end{document}